\documentclass[aps,prd,reprint,onecolumn,superscriptaddress]{revtex4-2}
\usepackage{graphicx}
\usepackage{amsmath}
\usepackage[dvipsnames]{xcolor}
\usepackage{upgreek}
\usepackage{lipsum}
\usepackage{calc}
\usepackage{array}
\usepackage{siunitx}
\usepackage{courier}
\DeclareSIUnit{\microsecond}{\SIUnitSymbolMicro s} 
\newcommand{\rev}[1]{\textcolor{black}{#1}}

\DeclareUnicodeCharacter{2009}{\,} 
\renewcommand{\selectlanguage}[1]{}

\usepackage{soul}
\usepackage[utf8]{inputenc}

\begin{document}
\title{Direct pulse-level compilation of arbitrary quantum logic gates on superconducting qutrits}

\author{Yujin Cho}
\email{cho25@llnl.gov}
\affiliation{Lawrence Livermore National Laboratory, Livermore, California 94550, USA}
\author{Kristin M. Beck}
\affiliation{Lawrence Livermore National Laboratory, Livermore, California 94550, USA}
\author{Alessandro R. Castelli}
\affiliation{Lawrence Livermore National Laboratory, Livermore, California 94550, USA}
\author{Kyle A. Wendt}
\affiliation{Lawrence Livermore National Laboratory, Livermore, California 94550, USA}
\author{Bram Evert}
\affiliation{Rigetti Computing, 775 Heinz Ave., Berkeley, California 94710, USA}
\author{Matthew J. Reagor}
\affiliation{Rigetti Computing, 775 Heinz Ave., Berkeley, California 94710, USA}
\author{Jonathan L DuBois}
\affiliation{Lawrence Livermore National Laboratory, Livermore, California 94550, USA}

\begin{abstract}
Advanced simulations and calculations on quantum computers require high-fidelity implementations of quantum operations. The universal gateset approach builds complex unitaries from a small set of primitive gates, often resulting in a long gate sequence which is typically a leading factor in the total accumulated error. Compiling a complex unitary for processors with higher-dimensional logical elements, such as qutrits, exacerbates the accumulated error per unitary, since an even longer gate sequence is required.
Optimal control methods promise time- and resource- efficient compact gate sequences and, therefore, higher fidelity. These methods generate pulses that can directly implement any complex unitary on a quantum device.
In this work, we demonstrate that any arbitrary qubit and qutrit gate can be realized with high-fidelity, which can significantly reduce the length of a gate sequence. We generate and test pulses for a large set of randomly selected arbitrary unitaries on several quantum processing units (QPUs): \rev{the LLNL Quantum Device and Integration Testbed (QuDIT)’s standard QPU and three of Rigetti's QPUs: Ankaa-2, Ankaa-9Q-1, and Aspen-M-3. On the QuDIT platform's standard QPU, the average fidelity of random \textit{qutrit} gates is $97.9\pm0.5\,\%$ measured with conventional QPT and $98.8\pm0.6\,\%$ from QPT with gate folding. Rigetti's Ankaa-2 achieves random \textit{qubit} gates with an average fidelity of $98.4\pm0.5\,\%$ (conventional QPT) and $99.7\pm0.1\,\%$ (QPT with gate folding). On Ankaa-9Q-1 and Aspen-M-3, the average fidelities with conventional qubit QPT measurements were higher than 99\,\% (see Appendix).}
We show that optimal control gates are robust to drift for at least three hours and that the same calibration parameters can be used for all implemented gates. Our work promises that the calibration overheads for optimal control gates can be made small enough to enable efficient quantum circuits based on this technique.
\end{abstract}
\maketitle

%%%%%%%%%%%%%%%%%%%%%%%%%%%%%%%%%%%%%%%%%%%%
\section{Introduction}
Superconductor-based quantum processors have improved significantly over the past few decades \cite{muller_towards_2019, kjaergaard_superconducting_2020}. Recently, several experiments have utilized quantum computers for simulations in quantum chemistry \cite{bennewitz_neural_2022, cao_quantum_2019, ryabinkin_qubit_2018} and quantum physics \cite{jafferis_traversable_2022, atas_su2_2021, atas_simulating_2023}. Although these small scale simulations show the potential of using quantum computers in large and complex simulations, current usage is limited by the coherence times of superconducting qubits and the errors accumulated during operations. 

Quantum circuits are, in general, constructed from primitive gates that have high fidelities. However, this often results in lengthy gate sequences and necessitates significant error mitigation \cite{cai_quantum_2023, bennewitz_neural_2022, endo_hybrid_2021}. The problem with long gate sequences is that even small errors in individual gates can accumulate and negatively impact the overall fidelity. Additionally, the coherence time of the QPU sets a limit on the total operation time. For instance, when simulating Hamiltonian dynamics in quantum systems, the evolution is typically represented using quantum logic gates. Instead of implementing the unitary using standard gates in a circuit model, an alternative approach involves generating pulses specifically designed for the desired unitary and directly implementing them in the hardware. This method reduces the number of gates required, resulting in faster operation of the quantum circuits with improved accuracy.

Compiling quantum circuits using standard gates becomes more challenging as the number of qubits or the complexity of the system increases, especially in the case of qudit systems. For example, the application of standard gates on one qubit can introduce phase errors in other qubits. These accumulated phase errors can be partially corrected by applying single-qubit $Z$ gates, but this correction is not universally applicable. In some cases, multi-qubit gates do not trivially commute with single-qubit $Z$ gates, requiring additional gates to correct such errors \cite{huang_quantum_2023}. Pulse-level compilation of quantum circuits can implicitly correct these phase errors, because the relative phase accumulation between different frequency frames is already taken into account in the Hamiltonian model \cite{koch_quantum_2022, wilhelm_introduction_2020}. This means that, in principle, the phase errors are eliminated or reduced to zero. 

For multi-level qudits, the problem of long gate sequences is exacerbated. Instead of a single qubit rotation, general qutrit gates decompose into a combination of $X$ and $Z$ gates with six standard $X_{\pi/2}$ gates in the $0-1$ and $1-2$ manifolds \cite{morvan_qutrit_2021}. In most quantum processors, virtual $Z$ gates are used, which makes the gate time practically zero. The total gate time by this decomposition is generally six times the duration of the $X_{\pi/2}$ gate.

Optimal control algorithms are utilized to find pulses that enable the direct implementation of custom gates at the pulse-level. Recent advancements have focused on developing efficient optimal control algorithms for both closed and open quantum systems \cite{goodwin_adaptive_2022, petersson_optimal_2022, petersson_discrete_2020, gunther_quandary_2021, chadwick_efficient_2023}. 
Experimental demonstrations have showcased the effectiveness of optimal control algorithms in various scenarios. These include single qubit gates \cite{werninghaus_leakage_2021, liebermann_optimal_2016}, the $0-2\,\textrm{SWAP}$ qutrit gate \cite{wu_high-fidelity_2020}, ququart gates \cite{seifert_exploring_2023}, and two-qubit gates \cite{howard_demonstrating_2022}. These demonstrations indicate that optimal control has the potential to enable universal control of quantum systems.
The application of pulse-level control facilitated by optimal control algorithms has shown promise in improving the fidelity of simulations. This has been demonstrated in small-scale problems in nuclear physics \cite{holland_optimal_2020} and plasma physics \cite{shi_simulating_2021}.
However, one challenge associated with optimal control is the requirement for extensive calibration to implement the obtained pulses on specific hardware architectures with high fidelity. This practical limitation has raised concerns about the feasibility of optimal control.

In this work, we present high-fidelity arbitrary quantum logic gates achieved by optimal control on two transmons that have different hardware architectures. Our results show that it is possible to prepare practically any random unitary with an average fidelity exceeding $99.7\,\%$ (qubit gates, Ankaa-2) and $98.8\,\%$ (qutrit gates, LLNL's QPU) while requiring minimal calibration. The required gate length for these random gates is shorter than the same operation implemented with primitive gate sets. Additionally, the simplified calibration process is easily transferrable to different hardware platforms.

\section{Method}
We tested optimal control pulses generated for random unitaries on four superconducting transmon quantum processors, the LLNL Quantum Device and Integration Testbed (QuDIT)'s standard QPU and Rigetti's \rev{Ankaa-2, Ankaa-9Q-1, and Aspen-M-3}. The QuDIT device has a single transmon made of tantalum on a sapphire substrate that has a long energy decay time \cite{place_new_2021, tennant_low-frequency_2022}. Rigetti's \rev{Ankaa-2 has 84 transmons}, and we choose one representative transmon among them to focus in this work. The hardware parameters on the two systems are listed in Table \ref{tab:param}. \rev{Additional data and their parameters from a single transmon on Ankaa-9Q-1 and Aspen-M-3 are presented in Appendix \ref{app:rigetti}.}

\begin{center}
\begin{table}[h]
\caption{Parameters for the LLNL QuDIT's standard QPU and the selected qubit on Rigetti's \rev{Ankaa-2}. $\omega_{ij}$ indicates the transition frequency from $|i\rangle$ to $|j\rangle$. $T_1^{ij}$ is the energy decay time and $T_2^{ij}$ is the decoherence time in $i-j$ manifold. The $T_2^*$ times were measured with Ramsey oscillation. \\}
\begin{tabular}{ c  r  r }
\hline
Parameters & LLNL QuDIT & Rigetti's \rev{Ankaa-2}\\ \hline
$\omega_{01}$ & 3.446 GHz & \rev{4.477 GHz} \\
$\omega_{12}$ & 3.237 GHz &  \rev{not measured}\\
$T_1^{01}$ & 220 $\si{\microsecond}$ & \rev{27} $\si{\microsecond}$\\
$T_2^{*,01}$ & 22 $\si{\microsecond}$ & \rev{21} $\si{\microsecond}$\\
$T_1^{12}$ & 145 $\si{\microsecond}$ & not measured \\
$T_2^{*,12}$ & 25 $\si{\microsecond}$ & not measured \\
\hline
\end{tabular}
\label{tab:param}
\end{table}
\end{center}

In a closed quantum system, the Hamiltonian $H$ of a superconducting transmon in the rotating frame is approximated by:
\begin{equation}
\label{eq:hamiltonian}
H= 0.5\alpha \,a^\dag a^\dag a a+ p(t)(a+a^\dag) + \textit{i}\,q(t)(a-a^\dag),
\end{equation}
up to $\mathcal{O}{(a^\dagger a)^2}$, where $\alpha=\omega_{12}-\omega_{01}$ is the anharmonicity, $a$ is the lowering operator, $p(t)$ and $q(t)$ are the control pulses given as time-dependent functions that we optimize. $\omega_{ij}$ indicates the transition frequency between $|i\rangle$ and $|j\rangle$.
For each target unitary, pulses were obtained either using a Julia open source package, JuQbox (for Rigetti's \rev{Ankaa-2}), or TensorOptimalControl (for LLNL QuDIT's device). JuQbox optimizes pulses that are parametrized with B-splines using L-BFGS algorithm \cite{petersson_optimal_2022, petersson_discrete_2020}. The B-splines provide amplitude modulation of harmonic signals at resonance frequencies. 
TensorOptimalControl is an LLNL developed GPU-accelerated suite for implementing gradient-based quantum-control fitting protocols.  This suite uses a short time Trotter expansion similar to the well-established GRAPE  \cite{khaneja_optimal_2005, machnes_comparing_2011} algorithm. The underlying TensorFlow library enables construction of exact gradients for arbitrary objective functions, and efficient use of GPUs allows for a run-time to scale at a log of the pulse length so long as the entire problem fits within the GPU memory.\rev{When compared to other open source GRAPE-like pulse fitting codes, TensorOptimalControl tends to run from 10$\times$ to 1000$\times$ faster depending on the available GPU for moderate gate sizes and pulses lengths.}
The typical time to solve for an arbitrary $SU(3)$ unitary with JuQbox or TensorOptimalControl is $\lesssim 15 \,\mathrm{s}$ on an Intel Core i5-9600K CPU. 

The pulse lengths were set to $220\,\mathrm{ns}$ for the QuDIT qutrit gates, and \rev{$40\,\mathrm{ns}$ for the Ankaa-2 qubit gates}. With these durations, we were able to achieve numerical convergence of the pulses with fidelity higher than $99.98\,\%$ for all tested gates. \rev{Any lower fidelity would result in poor fidelities in experiments.} The amplitudes of the pulses remain below the hardware limit of the arbitrary waveform generators on the two hardware platforms. 

\rev{The pulses were calculated with a resolution of 64 points per nanosecond with TensorOptimalControl and finer time-stepping with JuQbox (20 points per 1/$\alpha$) to accurately simulate the dynamics of the quantum system for the target gate operation. For qubit gates on Ankaa-2, since $\omega_{12}$ was not measured, we used $\alpha = 200\,\mathrm{MHz}$ which is a typical anharmonicity in a transmon. In this case, this value is used only for an optimization purpose and a small variation in $\alpha$ does not affect the fidelities of the gates. 
Subsequently, we perform downsampling on the pulses to achieve a rate of 1 point per nanosecond for hardware implementation. This is achieved by selecting 1 point out of every 64 data points. It is important to note that the control pulses do not contain frequency components that exceed 1 GHz, ensuring that no information is lost during the downsampling process. We also examined two alternative downsampling methods, namely averaging every 64 data points and decimation. However, our experimental observations revealed that the impact of downsampling methods on gate performance is smaller than what can be observed in the experiment.}

To achieve the best optimal control gate performance, we calibrate the pulse amplitudes using two scalar factors, $\gamma$ and $\sigma$.  
The calibrated pulses, $\mathcal{C}$, are written as:
\begin{equation}
\mathcal{C}(g(f)) = \gamma\big[ \mathcal{X}(f < \omega_{c}) + \sigma \cdot \mathcal{X}(f > \omega_{c}) \big],
\label{eq:pulse_calib}
\end{equation}
where $g(f)$ is the frequency domain representation of the control pulses, $p(t)$ and $q(t)$, $\mathcal{X}(\Delta f)$ is the spectral component in the frequency range $\Delta f$, and $\omega_c$ is the average frequency of $\omega_{01}$ and $\omega_{12}$.  
The amplitude scaling constant, $\gamma$, converts pulses calculated in units of the frequency ($\mathrm{Hz}$) to the physically applied voltage by the arbitrary waveform generators. To fine-tune $\gamma$, we measure state populations after performing a gate one to ten times, compare them to the predicted trajectories from the Lindblad master equation implemented in Python QuTiP package \cite{johansson_qutip_2012, johansson_qutip_2013}, and update $\gamma$ to minimize the difference between the measured and the predicted evolution. This process allows us to optimize $\gamma$ for any gate. 
By adjusting the spectral weight $\sigma$ between the $\omega_{01}$ and $\omega_{12}$ components, we compensate for any frequency dependence in the signal chain from the room temperature electronics to the device at 10 mK. This ratio also includes a factor of about $1.4$ from the lowering and the raising operators between different energy levels in the transmon Hamiltonian in the charge-basis \cite{wu_high-fidelity_2020}.
Previously, the weight calibration was done by constructing a densely-sampled spectral filter $\sigma(f)$ around the transition frequencies \cite{wu_high-fidelity_2020}, which could take longer than 10 minutes to measure. In all gates we sampled, the highest spectral components of the pulses were at $\omega_{01}$ and $\omega_{12}$, enabling us to instead perform a two-point calibration with a constant weight, $\sigma$, applied to the two transition frequencies. 

In the calibration process, we used a square-root of $0-2\,swap$ gate, $sSW02$, as a reference for qutrit gates. After applying it to the ground state, this gate generates an equal superposition of the states $|0\rangle$ and $|2\rangle$, and two applications of $sSW02$ realizes one $0-2\,swap$ gate. Using this gate, we calibrate $\gamma$ and $\sigma$ iteratively, until we find the optimal gate performance. 
In every measurement, we corrected the state-preparation-and-measurement (SPAM) error with an independently-measured confusion matrix. \rev{The average SPAM error for both systems is given in Appendix \ref{app:cmat}.} The best calibration parameters for  $sSW02$ were then applied to all random gates tested in this work. For qubit gates, we used $RX(\pi/2)$ pulse as a reference and adjusted the amplitude $\gamma$ to find the optimal calibration parameters. We tested 300 random qutrit gates on LLNL QuDIT, and a smaller sample of \rev{100 random qubit gates on Rigetti's Ankaa-2}.

%%%%%%%%%%%%%
%%% Tomography %%%
%%%%%%%%%%%%%
We evaluate the fidelities of the randomly generated optimal control gates using quantum process tomography (QPT).
For qutrit gates, we first prepare one of these nine different initial states \cite{chuang_prescription_1997}:
\begin{equation}
\begin{gathered}
|0\rangle, |1\rangle, |2\rangle \\
(|0\rangle + |1\rangle)/\sqrt{2}, (|1\rangle + |2\rangle)/\sqrt{2}, (|0\rangle + |2\rangle)/\sqrt{2} \\
(|0\rangle+i |1\rangle)/\sqrt{2}, (|1\rangle+i |2\rangle)/\sqrt{2}, (|0\rangle+i |2\rangle)/\sqrt{2}
\label{eqn:tomo_init}
\end{gathered}
\end{equation}
We then apply the gate to be evaluated, followed by a projection operator generated by one of the Gell-Mann matrices \cite{gell-mann_symmetries_1962}. For a given Gell-Mann matrix, $\lambda_m$, the corresponding operator is 
\begin{equation}
A_m = \mathrm{exp}(-i \pi/4 \lambda_m)\,.
\label{eq:complete_set}
\end{equation}

The process map, $\epsilon$, is defined as 
\begin{equation}
\rho_{out} = \epsilon ( \rho_{in} ) = \sum_{mn}A_m\rho_{in} A_n^\dagger \chi_{mn}
\label{eqn:tomo_map}
\end{equation}
where $\rho_{in}$ ($\rho_{out}$) is the input (output) density matrix and $\chi_{mn}$ is the $(m,n)$ element in the process matrix $\chi$. To find the process matrix from the measured data sets, we use the \textit{pgdB} algorithm, which is based on a gradient-descent optimization \cite{knee_quantum_2018}. The fidelity between the ideal, $\chi_{ideal}$, and the measured, $\chi_{meas}$, process matrices is calculated as trace-norm of $\chi_{ideal}^\dagger\chi_{meas}$. 
The process matrices are then mapped onto generalized Pauli basis, $P$  \cite{gottesman_fault-tolerant_1999}:
\begin{equation}
\begin{split}
P = \{&I, X_3^1, X_3^2, Z_3^1, Z_3^1X_3^1, \\
&Z_3^1X_3^2, Z_3^2, Z_3^2X_3^1, Z_3^2X_3^2\} 
\end{split}
\end{equation}
where $X_3$ and $Z_3$ are:
\[
%\begin{equation}
X_3=
\begin{pmatrix}
0 & 0 & 1\\
1 & 0  & 0\\
0 & 1 & 0 
\end{pmatrix}
%\end{equation}
,
Z_3=
\begin{pmatrix}
1 & 0 & 0\\
0 & \beta  & 0\\
0 & 0 & \beta^2 
\end{pmatrix},
\]
and $\beta$ is $-0.5+i\sqrt{3}/2$. 

To perform QPT measurements for a qubit gate in its $0-1$ frame, the initial state is prepared in one of these six states:
\begin{equation}
\begin{gathered}
|0\rangle, |1\rangle, (|0\rangle \pm |1\rangle)/\sqrt{2}, (|0\rangle\pm i |1\rangle)/\sqrt{2}
\label{eqn:tomo_init_2level}
\end{gathered}
\end{equation}
and the projection operators to project the Bloch vector to one of the Pauli $X$, $Y$, and $Z$ basis. 

%%%%%%%%%%
%%% Result %%%
%%%%%%%%%%
\section{Result}

The $sSW02$ unitary, a reference for arbitrary qutrit gates, is:
\begin{equation}
sSW02=
\begin{pmatrix}
1/\sqrt{2} & 0 & -i/\sqrt{2}\\
0 & 1  & 0\\
-i/\sqrt{2} & 0 & 1/\sqrt{2} 
\label{eq:sqswap02}
\end{pmatrix}
\end{equation}

%%%%%%%%%%%
%%%% Figure 1 %%%
%%%%%%%%%%%
\begin{center}
\begin{figure}[b]
\includegraphics[scale=1.0]{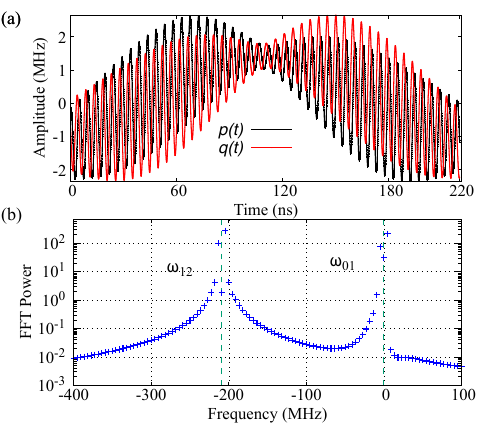}
\caption{(a) Control pulses in the qubit $0-1$ rotating frame for $sSW02$ on the QuDIT device: $p(t)$(black) represents the in-phase component and $q(t)$(red) shows the out-of-phase component. (b) Fast-fourier transform (FFT) of the pulse in panel (a) is shown. The frequency components are concentrated at the qubit $0-1$ and $1-2$ transition frequencies.}
\label{fig:pulse}
\end{figure}
\end{center}

Figure \ref{fig:pulse} shows the generated pulses for a $sSW02$ gate and its fast-fourier transform (FFT). The two pulses, $p(t)$ and $q(t)$, are the in- and out-of-phase components of the pulses in the qubit rotating frame. The frequency components of this pulse are concentrated at 0 and near $-210\,\mathrm{MHz}$ which correspond to $\omega_{01}$ and $\omega_{12}$, respectively. All the pulses generated for the random gates in this work have similar FFT profiles.

After generating these pulses, we calibrate them by adjusting the amplitude, $\gamma$, and the spectral weight, $\sigma$, in Eq. (\ref{eq:pulse_calib}), as described in Method section. The optimized $\sigma$ for $sSW02$ from this procedure is 1.8.

%%%%%%%%%%%
%%% Figure 2 %%%
%%%%%%%%%%%
\begin{figure*}
\includegraphics[scale=1.0]{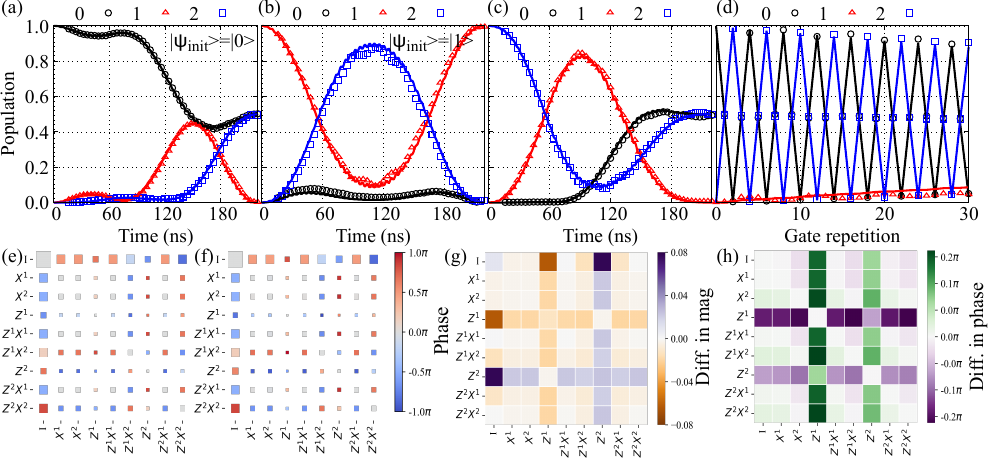}
\caption{(a-c) Time evolution of the quantum states when $sSW02$ is being played. Panel (a), (b), and (c) correspond to different initial states at $|0\rangle$, $|1\rangle$, and $|2\rangle$, respectively. (d) Populations on the three states change as the $sSW02$ gate is applied repeatedly up to 30 times. \rev{(a-d) The open symbols represent the measured data, and the solid lines depict the expected trajectories from the Lindblad master equation at each pulse time step (64 data points per second, panels a-c) and at integer numbers of gate repetition (panel d). In panel (d), we have connected the simulated data points at each repetition with solid lines to aid visualization.} Black open circles indicate the population on $|0\rangle$, red open triangles for $|1\rangle$, and blue open squares for $|2\rangle$. \rev{The reference and the measured process matrix of the $sSW02$ gate are shown in panel (e-f), respectively. The difference in magnitude and in phase between the two are presented in panel (g-h)}.}
\label{fig:sqswap02_exp}
\end{figure*}

To confirm our calibration parameters, we demonstrate the $sSW02$ gate performance with three different measurements \rev{as shown in Fig. \ref{fig:sqswap02_exp}: 1) time-evolution of quantum states during one gate application (panel a-c), 2) repeating the gate (panel d), and 3) QPT (panel e-h).} In Fig. \ref{fig:sqswap02_exp}(a-d), the expected trajectories (solid lines) are simulated \rev{with the Lindblad master equation whose Lindblad operators are constructed from $T_1$ and $T_2$.} To observe the time dynamics of population in each state during the gate application, we prepare the initial state to $|0\rangle$, $|1\rangle$, and $|2\rangle$ and measure the state populations every 4 ns. Figure \ref{fig:sqswap02_exp}(a-c) shows that the measurements agree with the expected ones within $\sim2\,\%$. Next, we apply the gate multiple times to amplify small errors that may not have been captured with fewer gate repetitions. Figure \ref{fig:sqswap02_exp}(d) shows an overall decay of the populations due to decoherence of the system, as expected from the simulations (solid lines). \rev{Coherent error in the operation is amplified as the gate is being repeatedly applied as shown in Appendix \ref{app:gate_rep}. }
%If the amplitude of the pulse is not calibrated to its optimal value, coherent error is amplified as we repeat the gate application, as shown in Appendix \ref{app:gate_rep}. Larger span of the gate repetition can also be found in Appendix \ref{app:gate_rep}.}
We finally perform QPT (see Method), to measure the fidelity of the $sSW02$ gate. Figure \ref{fig:sqswap02_exp}(e-f) shows the expected (panel e) and the measured (panel f) process matrices. The size and the color of the squares indicate the magnitudes and phases of each component in a process matrix, respectively. \rev{To illustrate the difference between the two process matrices, we plot the difference in magnitude and phase in Fig. \ref{fig:sqswap02_exp}(g-h). These plots show that there are small errors in the $Z^1$ and $Z^2$ bases\textemdash less than 0.08 in magnitude and $0.2\,\pi$ in phase.} The measured fidelity is $99.7\,\%$ ($96.5\,\%$), after (before) correcting the SPAM error.

%%%%%%%%%%
%%% Figure 3 %%%
%%%%%%%%%%
\begin{center}
\begin{figure}
\includegraphics[scale=1.0]{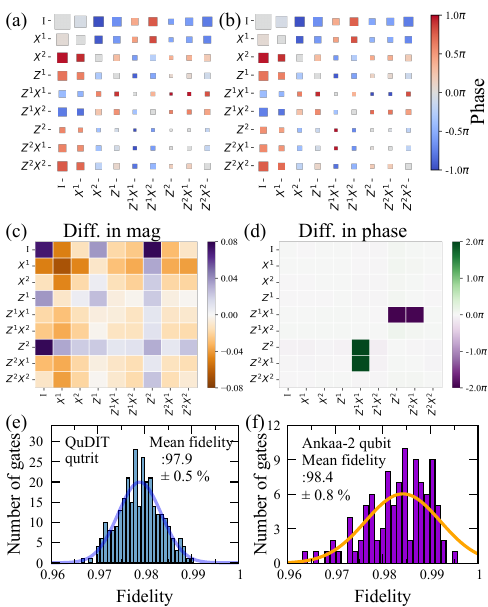}
\caption{(a) The reference and (b) the measured process matrices of a random gate.\rev{The difference between the two matrices is shown in panel (c-d) for magnitude and phase, repsectively.} (e) The average fidelity of 300 random qutrit gates on the LLNL's QuDIT platform is $97.9\pm0.5\%$. (d) The average fidelity of 100 random \textit{qubit} gates on \rev{Ankaa-2  is $98.4\pm0.8\%$}. 
}
\label{fig:random_gates}
\end{figure}
\end{center}

%To experimentally demonstrate that 
A simplified calibration procedure can guarantee a high fidelity for any random gate. We fix $\gamma$ and $\sigma$ to the calibrated values obtained from $sSW02$, and apply them to the pulses for 300 randomly generated qutrit gates, which were generated from a function, \texttt{rand\_unitary}, in the QuTiP python package \cite{johansson_qutip_2012, johansson_qutip_2013}. Figure \ref{fig:random_gates} shows measurements of an example randomly generated qutrit gate, $G$, whose unitary is given by: 
\begin{equation*}
G=
\begin{pmatrix}
0.599+0.194i & 0.673+0.026i & -0.387+0.020i\\
-0.654+0.163i & 0.638-0.335i  & 0.155-0.054i\\
0.387+0.027i & 0.164+0.011i & 0.887-0.191i 
\end{pmatrix}.
\end{equation*}

\rev{Figure \ref{fig:random_gates}(a-b) shows the reference and the measured process matrices. The magnitude error is less than $0.081$ as shown in Fig. \ref{fig:random_gates}(c). From the phase map presented in Fig. \ref{fig:random_gates}(d), we found out large phase errors on $Z^1X^1$ and $Z^2$ basis. The average amount of phase error is $0.07\pi$. By comparing the measured and the reference matrix, we obtained $99.0\,\%$ fidelity of the gate $G$.}
%The measured process matrix (panel (b)) looks very close to the ideal one (panel (a)) with small discrepancies, such as flipped phases in $Z^2-Z^1X^1$ and $Z^2X^1-Z^2X^1$ basis, indicating a small phase error induced by the gate. The obtained fidelity of the gate $G$ is $99.0\,\%$. 
We repeat the same analysis for 299 different random gates on the QuDIT platform. The mean fidelity is $97.9\pm0.5\,\%$ while the highest fidelity is $99.0\,\%$, as presented in Fig. \ref{fig:random_gates}(e). 

This calibration procedure can easily be transferred to a different hardware architecture. To demonstrate this, we follow the same calibration procedure and measure the qubit random gate fidelities on Rigetti's \rev{Ankaa-2}. For 100 randomly generated qubit gates, the highest and lowest fidelities are $99.9\,\%$ and $96.4\,\%$, respectively. \rev{In addition to our measurement on Rigetti's Ankaa-2, we performed optimal control on arbitrary qubit gates on Rigetti's Ankaa-9Q-1 and Aspen-M-3 (see Appendix \ref{app:rigetti}). The average fidelities are $99.1\pm0.4\,\%$ for Ankaa-9Q-1 and $99.5\pm0.3\,\%$ for Aspen-M-3.} 
%When we measured the fidelity of random gates performed on the three levels on the same transmon, we achieved the average fidelity of $85.6\pm3.6\,\%$, after correcting the SPAM error (see Discussion).
%We attribute the low-fidelity cases to under-sampling. For short pulses, such as our 40 ns pulses, when the generated pulses vary too rapidly, down-sampling to 1 point per nanosecond could overly simplify the spectral features, resulting in a low fidelity. The average fidelity of the measured gates, excluding such under-sampled cases, is $99.5\pm0.3\,\%$.

%and qutrit random gate fidelities on Rigetti's Aspen-M-3. For 54 random gates on qubit $0-1$ frame, the highest and lowest fidelities are $99.9\,\%$ and $89.8\,\%$, respectively. We attribute the low-fidelity cases to under-sampling. For short pulses, such as our 40 ns pulses, when the generated pulses vary too rapidly, down-sampling to 1 point per nanosecond could overly simplify the spectral features, resulting in a low fidelity. The average fidelity of the measured gates, excluding such under-sampled cases, is $99.5\pm0.3\,\%$. When we measured the fidelity of random gates performed on the three levels on the same transmon, we achieved the average fidelity of $85.6\pm3.6\,\%$, after correcting the SPAM error (see Discussion).

%%%%%%%%%%%
%%% Figure 4 %%%
%%%%%%%%%%%
\begin{center}
\begin{figure}[h]
\includegraphics[scale=1.0]{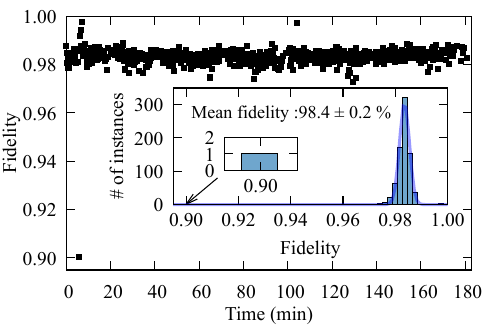}
\caption{Fidelity of $sSW02$ gate changes over time. The mean fidelity is $98.4\,\%$ while the highest fidelity is $99.6\,\%$.}
\label{fig:time_dep}
\end{figure}
\end{center}

%In Fig. \ref{fig:random_gates} and Appendix \ref{app:rigetti}, 
In the measurements on the tested QPUs, the standard deviations of the fidelities are between $0.3\,\%$ and $0.8\,\%$. One possible reason is temporal variations such as fluctuations of the quantum system over time. To measure the temporal fluctuation of fidelity, we monitor the fidelity of the $sSW02$ gate for 3 hours on the QuDIT platform, which is comparable to the duration of the full data set plotted in Fig. \ref{fig:random_gates}(e). Immediately after the calibration, the measured fidelity was $99.7\,\%$, followed by repeated measurements every $13-14$ seconds, resulting in 782 measurements. Figure \ref{fig:time_dep} shows that the mean fidelity is $98.4\pm0.2\,\%$ with a fidelity drop at one point, as low as $90\,\%$. This low-fidelity is quickly recovered to $98.6\,\%$ on the next measurement. The highest fidelity measured is $99.6\,\%$. This $0.2\,\%$ fluctuation in the fidelity of a single gate over time partially accounts for the $0.5\,\%$ spread in gate fidelities in our random sample (Fig. \ref{fig:random_gates}(c)). \rev{We will discuss the cause of temporal fluctuation and other potential reason for the fidelity span later in Discussion.}
%We repeat this measurement on Aspen-M-3. We monitor the fluctuation of the fidelity of one random $qubit$ gate in 50 separate measurements over the course of 15 minutes. The fidelity fluctuation is $0.1\,\%$, which should be compared with the $0.3\,\%$ spread in the corresponding random gate sample. 

%%%%%%%%%%%
%%% Figure 5 %%%
%%%%%%%%%%%
\begin{center}
\begin{figure}[b]
\includegraphics[scale=1.0]{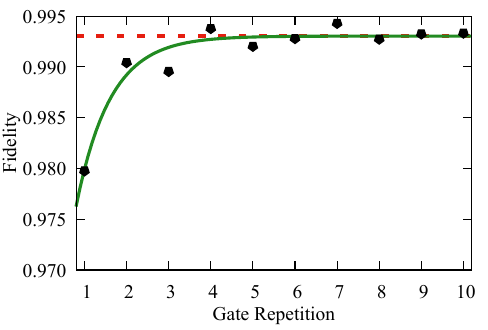}
\caption{The fidelity of a single random gate was extracted by performing QPT with 1 to 10 applications of the same gate with the spectral weight $\sigma=1.87$. The fidelity converges to $99.3\,\%$. The green line is a guideline to visualize the convergence of the fidelities.}
\label{fig:grep_fid}
\end{figure}
\end{center}

\rev{QPT provides a thorough information about a gate process. However, QPT is not immue to certain sources of error, such as the SPAM error and the infidelities of operators used for state preparation and projection. Firstly, we correct for the SPAM error by applying a confusion matrix. The average confusion matrices are given in Appendix \ref{app:cmat}. Secondly, we mitigate the operator infidelities by measuring QPT with gate folding (gQPT). By applying the test gate repeatedly, we selectively amplify errors only in the gate itself, while keeping the errors in the operators unchanged. This allows us to isolate and quantify the errors specifically associated with the gate under test. By doing so, we can accurately assess the performance of the gate and identify potential area for improvement.}
%Another contribution to the measured gate fidelities is that the fidelities we measure with QPT are, in fact, the convolution of the pure gate fidelity and the state preparation and projection operator fidelities.
%To mitigate operator infidelity, we apply a random gate, $K$, repeatedly 1-10 times between the initial and the projection operators, performing QPT on $K^n$ at $n$-times gate folding, \rev{which we call gQPT}. 
\rev{Assuming that errors in each gate operation remains unchanged during the measurement and the operation time is short enough to be insensitive to decoherence, the fidelity can be evaluated by $1/n$-th root of the measured fidelity at $n$-th gate folding.} \rev{In other words, fidelity $\mathcal{F}$ is measured at $n$-th repetition of gate $K$, which absorbs the fidelity of the preparation and projection operators, and the adjusted fidelity is $\mathcal{F}^{1/n}$.} 

Figure \ref{fig:grep_fid} shows the extracted fidelity of one random gate as a function of the number of gate repetitions. \rev{The final gate fidelity is obtained from the exponential fit (green curve), given by $a\cdot \mathrm{exp}(-n/b) + c$ where $a$, $b$, and $c$ are the fitting parameters. The parameter $c$ corresponds to the converged fidelity that we extract.} \rev{We repeat gQPT for 8 different random qutrit gates whose initial fidelity ($n=1$) is $97.9\pm0.3\,\%$ on average,} and overall, the gate fidelity improves and converges to a higher fidelity around $99\,\%$. At least $1\,\%$ of the gate fidelities measured on the QuDIT QPU can be accounted for by the infidelity of the state preparation and projection operators. \rev{Similarly, we performed gQPT on 15 randomly selected qubit gates on Rigetti's Ankaa-2 (initial average fidelity: $97.3\pm0.6\,\%$) and achieved average fidelity $99.7\pm0.1\,\%$. To compare gQPT to other existing characterization protocols, we measured the fidelities of these 15 random gates using cross-entropy benchmarking (XEB), which gives us an average fidelity of $98.9\pm0.6\,\%$. The data can be found in Appendix \ref{app:benchmarking}}.

%%%%%%%%%%%%
%%% Discussion %%%
%%%%%%%%%%%%

\section{Discussion}
Optimal control can be an efficient tool to implement any high fidelity gates, as we present in this work. Our work shows that we can achieve $98-99\,\%$ average gate fidelity for randomly-generated gates with fast one-time calibration that can be applied to any random gate we want to produce. 

The fidelity measured with QPT is a lower bound of the measured gate fidelity. 
\rev{To measure true fidelities of test gates apart from SPAM error and operator infidelities, we selectively amplified errors in the test gate by gate folding and extracted the fidelity from exponential fit. This method is valid in the presence of coherent errors as long as the circuit at a large gate repetition is short compared to decoherence times of the device \cite{giurgica-tiron_digital_2020}. However, QPT with gate folding does not capture other types of errors, such as statistical errors (shot noise) which typically follows a Gaussian distribution.}
\rev{The extracted gate fidelity from gQPT increases by about $1\,\%$ over conventional QPT.} \rev{Comparing to XEB, gQPT estimates the fidelities to be about $0.8\,\%$ higher than XEB measurements. We attributes the difference to the operator infidelities in XEB measurements. XEB is inherently free from measurement errors, but not from operator infidelities. Single qubit primitive gates can be calibrated to very high fidelities, but small infidelities can accumulate at each application, which can contribute to slightly lower fidelities of XEB measurements.} 

\rev{During the experiments, we have observed fluctuation of a gate fidelity over time as shown in Fig. \ref{fig:time_dep}. Quantum systems, in general, exhibit fluctuations and drift over time, which can be attributed to various noise sources. For instance, inherent defects in quantum devices, such as residues from the fabrication process or adsorbents from the air, can cause resonant coupling with the qubits, leading to shifts in qubit frequencies and a sharp decrease in $T_1$ in the order of a few minutes \cite{klimov_fluctuations_2018}. Additionally, cosmic rays can generate quasiparticles that temporarily reduce $T_1$, which affects the quantum system on the order of milliseconds. Interaction with the environment can cause a long-term drift over a few hours. Due to the need for control, quantum devices cannot be perfectly isolated from the environment. Even though the coupling is weak, small environmental changes, like fluctuations in room temperature, can introduce noise into the device. Similarly, instability in room temperature electronics can introduce control errors. For instance, the amplitude of a $\pi$ pulse need occasional recalibration to ensure optimal performance. Reducing noise sources and mitigating their impact on qubit performance is an active area of research.}
The calibration parameters are stable for several hours and require a fine-tuning of the amplitude at the $1\,\%$-level after two weeks due to drift in the lab environment and the quantum system. 
%Firstly, due to the need for control, quantum devices cannot be perfectly isolated from the environment. Even though the coupling is weak, small environmental changes, like fluctuations in room temperature, can introduce noise into the device. Secondly, inherent defects in quantum devices, such as residues from the fabrication process or adsorbents from the air, can cause resonant coupling with the qubits, leading to shifts in qubit frequencies and a sharp decrease in $T_1$ \cite{klimov_fluctuations_2018}. Additionally, cosmic rays can generate quasiparticles that temporarily reduce $T_1$. Lastly, instability in room temperature electronics can introduce control errors. For instance, the amplitude of a $\pi$ pulse need occasional recalibration to ensure optimal performance. Reducing noise sources and mitigating their impact on qubit performance is an active area of research.}

\rev{In addition, the standard deviations of the measured fidelities are $0.3-0.8\,\%$, which is larger than expected from temporal fluctuation. This fidelity range could be related to the choice of universal parameters. Different qutrit gates have different ratio of the spectral components between the $0-1$ and $1-2$ transitions, which could lead to slightly adjusted spectral weight $\sigma$. For example, we optimized the spectral weight for the specific gate we present in Fig. \ref{fig:grep_fid}. For this gate, the highest fidelity was at $\sigma=1.87$, instead of the $\sigma=1.8$ for the reference gate. This suggests that adding a parameter that depends on the weight of spectral component may help implementing higher gate fidelities with narrower standard deviation. On the other hand, arbitrary qubit gates are independent of $\sigma$, because the pulse has only one frequency component. In this case, the fidelities of arbitrary qubit gates could be improved by frequent tuning of the amplitude scaling constant $\gamma$. Quantum hardware requires regular tuning of quantum gates to ensure its best performance by optimizing pulse amplitude. If the amplitude from control electronics is unstable, the system would require more frequent tuning of the pulse amplitude. }
%In all measurements, we observed that the standard deviations of the measured fidelities are $0.3-0.8\,\%$. These deviations may be related to the temporal variation of a gate as discussed earlier and shown in Fig. \ref{fig:time_dep}, and more importantly, choice of the universal parameters. 
%Different random gates have different ratio of the spectral components between the $0-1$ and $1-2$ transitions, which could lead to slightly adjusted spectral weight $\sigma$. For example, we optimized the spectral weight for the specific gate we present in Fig. \ref{fig:grep_fid}. For this gate, the highest fidelity was at $\sigma=1.87$, instead of the $\sigma=1.8$ for the reference gate. This suggests that adding a parameter that depends on the weight of spectral component may help implementing higher gate fidelities with narrower standard deviation.}

\rev{We have tested arbitrary qutrit gates using optimal control on Rigetti's Aspen-M-3 device. However, we achieved only $85.6\pm3.6\,\%$, which is more than $10\,\%$ lower than what we measured on LLNL's QPU.  This result indicates that the performance of arbitrary qutrit gates implemented with optimal control is system-dependent. Multi-qubit chips typically exhibit higher levels of noise compared to single qubit devices, primarily due to additional noise sources like flux or interactions between qubits. These noise sources are not static and can change over time.
In order to determine the frequencies required for the 0-1 and 1-2 transitions in the Hamiltonian model used for optimal control, we relied on Rigetti's calibration data, which is regularly updated every 6 hours. However, if the qubit frequencies experience significant drift within this 6-hour timeframe, the accuracy of the Hamiltonian model decreases, leading to phase errors accumulating during pulse execution.
Towards the end of the 6-hour calibration block, we noticed increasing coherent errors when using the Rigetti's pre-defined native gates. This suggests that the amplitude factor for the optimal control pulses also needs to be recalibrated.
Unfortunately, due to limited resources and accessibility to the Rigetti device, we were unable to thoroughly investigate all potential sources of error.}
%A fruitful direction is to develop a noise-resilient model for optimal control that takes into account common noise sources.}

%There is a small loss in fidelity by using universal parameters. Physically, different random gates have different ratio of the spectral components between the $0-1$ and $1-2$ transitions, which could lead to slightly adjusted spectral weight $\sigma$. For example, we optimized the spectral weight for the specific gate we present in Fig. \ref{fig:gate_rep}. For this gate, the highest fidelity was at $\sigma=1.87$, instead of the $\sigma=1.8$ optimized for $sSW02$. 

%% Future direction
One future direction is to apply this method to more robust optimal control gates. When we monitor the fidelity of one gate over a few hours, the fidelity drops by $8\,\%$ at one time and the standard deviation is $0.2\,\%$ as presented in Fig. \ref{fig:time_dep}. To generate robust pulses for a target unitary over time that is more stable, it would be useful to build a model that captures the time-dependence of the system, such as $T_1$, $T_2$, and qubit frequency fluctuation that is typically in the order of a few $\mathrm{kHz}$ \cite{peng_deterministic_2023}. 
Another direction would be to explore systematic way to achieve fast control of an arbitrary gate. To achieve the shortest gate time, there are a few challenges to overcome. When a gate becomes shorter, it tends to have higher amplitudes, which can unintentionally drive higher energy excitation. In addition, the pulse length is often limited by the clock cycle of the arbitrary waveform generator, typically 4 ns, limiting our ability to fully explore the dynamic range of the pulse length. \rev{Lastly, we plan to expand this calibration method to multi qubit and qutrit entangling gates to achieve high performance entangling operations.}

\section{Conclusion}
In this work, we experimentally demonstrate that optimal control technique can prepare any random quantum logic gate with minimal calibration at high fidelities, opening the door to greater adaptation of this technique. Our calibration procedure is applicable to different hardware architectures, showing that the optimal control is a practical and promising direction for optimized quantum circuits. 

Implementing custom gates at the pulse-level enables us to operate quantum simulations and algorithms faster with higher fidelities. For example, in Quantum Fourier Transform \cite{nielsen_quantum_2010, weinstein_implementation_2001} or variational quantum eigensolver (VQE) \cite{peruzzo_variational_2014,kandala_hardware-efficient_2017,wang_accelerated_2019, tilly_variational_2022}, a sequence of fixed gates can be replaced with an optimal control pulse to reduce the operation time and the overall gate count in the circuit. Similar ideas have been suggested to use parametrized pulses as an ansatz for higher fidelity VQE calculation \cite{liang_pan_2022}.

\section{Acknowledgement}
This work was supported by U.S. Department of Energy under grant number SC-FES SCW1736.
This work was performed under the auspices of the U.S. Department of Energy by Lawrence Livermore National Laboratory under Contract DE-AC52-07NA27344.

\appendix
%%%%%%%%%%%%%%%%%%%%
\rev{
\section{Performance of arbitrary gates on different hardware}
\label{app:rigetti}
Figure \ref{fig:rigetti} shows histograms of fidelities of arbitrary gates on Ankaa-9Q-1 and Aspen-M-3, whose average fidelities are $99.1\pm0.4\,\%$ and $99.5\pm0.3\,\%$, respectively. The hardware parameters are given in Table \ref{tab:param2}.
\begin{figure}[h]
\includegraphics{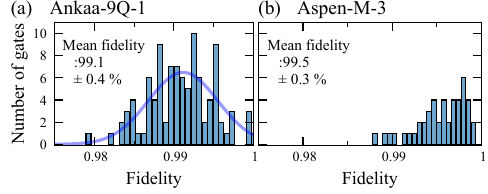}
\caption{Histogram of fidelities of (a) 100 randomly generated qubit gates on Ankaa-9Q-1 and (b) 46 randomly generated gates on Aspen-M-3.}
\label{fig:rigetti}
\end{figure}
\begin{center}
\begin{table}[h]
\rev{
\begin{tabular}{ c  r  r }
\hline
Parameters & Ankaa-9Q-1 & Aspen-M-3\\ \hline
$\omega_{01}$ & 4.526 GHz & 3.883 GHz \\
$T_1^{01}$ & 17 $\si{\microsecond}$ & 22 $\si{\microsecond}$\\
$T_2^{*,01}$ & 16 $\si{\microsecond}$ & 42 $\si{\microsecond}$\\
\hline
\end{tabular}
\caption{Parameters for a qubit on Rigetti's Ankaa-9Q-1 and Aspen-M-3. $\omega_{ij}$ indicates the transition frequency from $|i\rangle$ to $|j\rangle$. $T_1^{ij}$ is the energy decay time and $T_2^{ij}$ is the decoherence time in $i-j$ manifold. The $T_2^*$ times were measured with Ramsey oscillation. \\}
\label{tab:param2}
  }
\end{table}
\end{center}
}
%%%%%%%%%%%%%%%%%%%%%%
% \section{Pulse fidelity vs. measured fidelity}
% \label{app:pulse_fid}
% We measured the pulse fidelity by comparing propagator generated by the optimized pulses and the one from the target unitary, and compared it to the measured fidelity, as shown in Fig. \ref{fig:pulse_fid}. 
% \begin{figure}[h]
% \includegraphics[scale=1]{./figures/Sfig4.pdf}
% \caption{We compare measured fidelities (black, left axis) and pulse fidelities (red, right axis). The blue vertical lines indicate the cases that show low fidelities.}
% \label{fig:pulse_fid}
% \end{figure}
%%%%%%%%%%%%%%%%%%%%%
\rev{
\section{Confusion matrices}
\label{app:cmat}
To mitigate state preparation and measurement (SPAM) errors, we measured a confusion matrix before each measurement. On average, the confusion matrices on each system is given by:
LLNL QuDIT's qutrit
\begin{equation*}
\begin{pmatrix}
0.99672 & 0.00303 & 0.00024 \\
0.03184 & 0.95166 & 0.01650 \\
0.01269 & 0.04198 & 0.94534 
\end{pmatrix}
\end{equation*}
and for qubit gates on Rigetti's Ankaa-2,
\begin{equation*}
\begin{pmatrix}
0.9905 & 0.0095  \\
0.04 & 0.96 
\end{pmatrix}
\end{equation*}
%%%%%%%%%%%%%%%%%%%%%
\section{Coherent and incoherent error in repeated gate measurement}
\label{app:gate_rep}
\begin{figure}[h]
\includegraphics{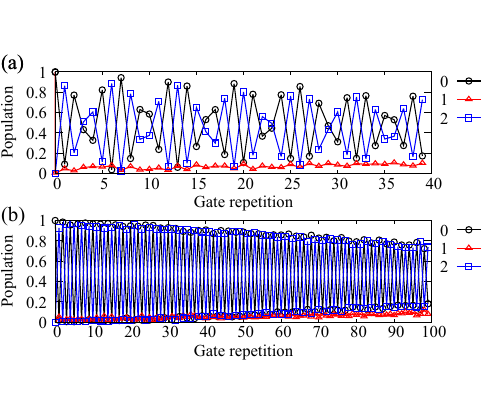}
\caption{\rev{Populations on the three levels when we repeatedly apply the gate: (a) before the amplitude of the pulses are calibrated and (b) after the calibration.}}
\label{fig:gate_rep}
\end{figure}
Coherent errors, resulting from miscalibrated pulse amplitudes, accumulate with each application of the gate, as shown in Fig. \ref{fig:gate_rep}(a). By precisely adjusting the pulse amplitude through multiple gate repetitions, we can mitigate the impact of coherent errors (see Fig. \ref{fig:gate_rep}(b)). When the gate is repeated up to 100 times, the overall population experiences decay due to the influence of $T_1$. Consequently, the coherent error becomes indiscernible in this scenario.
%%%%%%%%%%%%%%%%%%%%%
\section{Comparison between fidelities measured with XEB and QPT}
\label{app:benchmarking}
We measured fidelities of arbitrary gates using a conventional QPT, gQPT, and cross-entropy benchmarking (XEB). We followed the same protocols for QPT and gQPT as described in the main text. 
In XEB, we increased the circuit depth up to 256 to extract the fidelities and further used exponential decay fit to extract the cycle fidelity marked as open-red triangles in Fig. \ref{fig:fid_metric}.
\begin{figure}[t]
\includegraphics{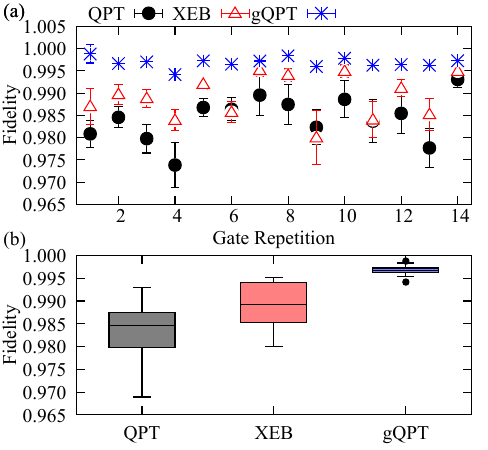}
\caption{For 14 randomly generated gates, we used a regular QPT (black), XEB (red), and QPT with repeated gates, named as \textit{gQPT} (blue).}
\label{fig:fid_metric}
\end{figure}
Figure \ref{fig:fid_metric} compares fidelities of 15 randomly generated qubit gates, measured with three methods on Rigetti's Ankaa-2. Average fidelities are $98.3\pm0.7\,\%$ (conventional QPT, black circles), $98.9\pm0.6\,\%$ (XEB, red triangles), and $99.7\pm0.1\,\%$ (gQPT, blue crosses). Overall gate fidelity measured by a conventional QPT with SPAM error mitigation matches with the one measured with XEB within uncertainties. gQPT extracts a higher average gate fidelity by almost $1\,\%$ than either QPT or XEB. QPT and XEB routines use primitive gates to construct the routines. Even though we mitigate SPAM error in QPT by applying a confusion matrix and XEB is inherently robust to SPAM errors, they are still sensitive to the operator fidelities. By repeating a gate between initial and measurement operators in gQPT, we can measure fidelities of a gate more accurately.
}

%\bibliography{Random_gate_library}

%

\end{document}